\documentclass[twocolumn,showpacs,aps,prb,amsmath,amssymb]{revtex4}

\usepackage{graphicx}

\begin{document}

\title{Domain wall entropy of the bimodal two-dimensional Ising spin glass}

\author{A. Aromsawa}
\affiliation{Department of Mathematics, Faculty of Science,
Mahidol University, Rama 6 Road, Bangkok 10400, Thailand}

\author{J. Poulter}
\affiliation{Department of Mathematics, Faculty of Science,
Mahidol University, Rama 6 Road, Bangkok 10400, Thailand}

\date{\today}

\begin{abstract}

We report calculations of the domain wall entropy for the bimodal two-dimensional
Ising spin glass in the critical ground state. The $L \times L$ system sizes
are large with $L$ up to $256$. We find that it is possible to fit the variance of
the domain wall entropy to a power function of $L$. However, the quality of the
data distributions are unsatisfactory with large $L > 96$. Consequently, it is
not possible to reliably determine the fractal dimension of the domain walls.

\end{abstract}

\pacs{75.10.Hk, 75.10.Nr, 75.40.Mg, 75.60.Ch}

\maketitle

\section{\label{sec:intro}Introduction}

The short-range Ising spin glass is still a source of controversy in spite of
its comparative simplicity. In brief, the exact mechanisms by which widely
separated spins influence each other are not clearly understood. The system
Hamiltonian, due to Edwards and Anderson \cite {EA75}, is

\begin{equation}
H=-\sum_{<ij>}J_{ij}\sigma_{i}\sigma_{j}
\label{e:EAH}
\end{equation}
where the nearest-neighbor exchange interactions $J_{ij}$ are
quenched random variables which, for present purposes, take random sign with
equal probability. The two particular models of disorder widely studied are
the bimodal, or $\pm J$, model where the interactions have fixed magnitude and
the Gaussian model where they are taken from a continuous normal distribution
of zero mean. In two dimensions it is generally accepted that the spin glass
only exists at zero temperature \cite {HY01,H01}, that is the critical
temperature $T_c=0$.

Spin correlations in two dimensions for bimodal disorder are expected to decay
algebraically according to

\begin{equation}
[<\sigma_{0}\sigma_{R}>^{2}]_{av}\sim R^{-\eta}
\label{e:SS1}
\end{equation}
in the ground state. To date, the best estimates from Monte Carlo simulation
\cite {KL04} and exact calculation \cite {PB05} seem to agree that the
exponent is $\eta = 0.14$. For Gaussian disorder, the exponent is
zero since the ground state, apart from a global inversion, is unique.
In contrast, bimodal disorder comes with a large ground state degeneracy
corresponding to an entropy of $0.07k$ per spin \cite{CM83,WS88,BP91,SK93,
BGP98,LGM04}. In spite of this it has been argued \cite {JLM06} that 
$\eta = 0$ in the thermodynamic limit. Very recently, Hartmann \cite {Hart07}
has estimated $\eta = 0.22$ from scaling arguments. This is in fair agreement
with a number of previous estimates \cite {WS88,Heidel,BY88,SK93,Ozeki}. 
Nevertheless, the important issue here is universality; that is whether or not
$\eta$ is positive.

Clear evidence of the importance of long range influence in the case of
bimodal disorder is apparent in the current controversy regarding the lowest
excited state. This issue dates back to the work of Wang and Swendsen
\cite {WS88} which proposed that the energy gap should be $2J$, not $4J$, in
the thermodynamic limit. Although some work \cite {SK93,KL04} has supported
the simple naive $4J$ scenario, it is now becoming clearer that strong
disorder is indeed producing a different result for the infinite system.
The energy gap has been often reported \cite {LGM04,W05,KLC05} as $2J$. Further, it has even
been suggested, based on a finite size scaling study of the correlation length
and the spin glass susceptibility \cite {JLM06}, that the specific heat should obey a
power law. This may indicate a feature universal with Gaussian disorder. 
A good review of the
issues involved here is given in Ref. \onlinecite {KLC07}. Essentially it
appears to be the case that there exist low energy excitations with very
long range influence.

A standard technique commonly used to investigate the long range response of
a spin glass is to introduce a domain wall defect \cite {Mc84,BM85,Heidel}. This is
done, in two dimensions, by drawing a line across the system and reversing the
signs of all bonds cut by the line. 
With Gaussian disorder, the ground and low energy excited states
are unique. Consequently, the term domain wall has a clear physical meaning and
appears as an optimium fractal path that corresponds to the lowest excited state.
The case of bimodal disorder is quite different as a result of the large degeneracy of
both the ground and excited states. The degeneracy of the ground state is of
the order of $\exp(0.07 L^2)$ for a $L \times L$ lattice and that of the first excited
state is probably much larger still \cite {LGM04}. In particular, it is often the case that the
defect system is not an excited state of the reference system. Thus, we do not have real
domain walls in the same sense as with the case of Gaussian disorder. 
Nevertheless, we might expect some useful knowledge to be derived from a study
of the bimodal system in response to reversing the signs of bonds along a line as
with the Gaussian system.

Thinking of the thermodynamic limit, we can discuss this issue in the light
of droplet theory \cite {FH88}. For the case of a continuous defect
distribution, Gaussian for instance, there is a unique ground state and we
define a droplet as a region bounded by a closed path (or surface in three
dimensions). Below the transition temperature, which is positive in three
dimensions at least \cite {KKY06}, the scaling properties of droplets of
various sizes can be investigated by reversing all spins inside the closed
path. This creates an excited state of the reference system and it is known
that there exist low energy droplet excitations of large spatial extent.
The idea here is
that a domain wall defect is much the same as a large droplet excitation
in the thermodynamic limit. For bimodal disorder this comparison is less 
clear due to the huge ground state degeneracy. Certain droplets may not
represent excitations at all. 

The most important prediction of droplet theory is that the energy difference
$E_{dw}$ between the two systems, with and without the domain wall, can be fitted
to a power function of the system size $L$. We have, for the spin glass,

\begin{equation}
<|E_{dw}|> \sim L^{\theta}
\label{e:Stiff}
\end{equation}
where $\theta$ is known as the spin-glass stiffness exponent. In three dimensions,
this exponent is found to be positive \cite {H99,PY00}, at about $0.2$ for both bimodal
and Gaussian disorder, showing consistency
with the existence of a stable spin glass phase at finite temperature. As a matter of
fact the droplet theory \cite {FH88} was originally developed for this type of case where both
the critical temperature and stiffness exponent are positive.
It is also probable that the Gaussian and bimodal models fit into the
same universality class \cite {KKY06} with respect to their transitions at finite
critical temperature. Degeneracy is not an issue due to the thermal fluctuations.

In two dimensions it was not immediately obvious that droplet theory \cite {FH88} is
entirely appropriate since the critical temperature is zero and the stiffness
exponent is not positive. For Gaussian disorder the stiffness exponent is
negative \cite {HY02}, $\theta \approx -0.28$, clearly indicating that the
spin glass is unstable at any finite temperature. It is also remarkable that the 
values of $\theta$ obtained from domain wall and droplet calculations are in very good 
agreement.
For bimodal disorder it is generally accepted that the stiffness exponent due to 
domain wall defects is zero \cite {HY01,AMM03,CHK04}. Nevertheless the situation 
for droplets is not clear. Hartmann \cite {Hart07} has estimated the droplet stiffness
exponent by constructing ground and first excited states of three models of bimodal
disorder and found good agreement with Gaussian disorder. Nevertheless, universality
is not shown since the correlation function exponent $\eta$ is not zero.

For the case of Gaussian disorder, in two dimensions, it has been suggested
\cite {AHH06,BDM06} that the domain walls are stochastic Loewner evolution
processes \cite {Cardy}. This theory is able to relate the domain wall
fractal dimension $d_f$ to the stiffness exponent via

\begin{equation}
d_f = 1 + \frac{3}{4(3+\theta)}
\label{e:SLE}
\end{equation}
and the result $d_f \approx 1.27$ agrees well with the literature
\cite {HY02}. There is also an interesting conjecture due to Fisch \cite {Conject}
that $\theta = (\surd 6 - 3)/2$ exactly. Nevertheless, there is no good reason to believe that
Eq. (\ref{e:SLE}) can be used with bimodal disorder, possibly due to the degeneracy
of the ground state.

The domain wall entropy $S_{dw}$ is defined in the same manner as $E_{dw}$.
Droplet theory \cite {FH88} predicts that $S_{dw}$ should take values with
random sign and large variance. In particular, the variance is predicted
to scale as

\begin{equation}
<S_{dw}^2> - <S_{dw}>^2 \sim L^{d_f}
\label{e:variance}
\end{equation}
which provides a possible means to estimate $d_f$. In particular, we can
use this to test the appropriateness of droplet theory for the case where
both the critical temperature and the stiffness exponent tend to zero from
above. This is precisely the situation we have for the two-dimensional model
with bimodal disorder.

Previous estimations of the fractal dimension from direct studies of the
domain wall entropy have been published. Saul and Kardar \cite {SK93} predict
$d_f \approx 1.0$. Fisch \cite {Fisch} argues that $d_f$ might be an
increasing function of $|E_{dw}|$. The possibility that $d_f = 1.25$ in
agreement with Eq. (\ref{e:SLE}) is not ruled out. Finally, Lukic
{\it et al} \cite {LMM06} have reported $d_f = 1.03(2)$. These values should
also be compared with those from topological analysis of the ground state.
Rom\'{a} {\it et al} \cite {RRR07} report $d_f = 1.30(1)$ while Melchert and
Hartmann \cite {MH07} find an interval $1.095(1) \leq d_f \leq 1.395(1)$.

In this article we report calculations of the domain wall entropy on
sample sizes that are much larger than anything done before. Furthermore,
our method is applied at an arbitrarily small temperature and there is
no need to extrapolate to the ground state. Our main conclusion is that the
domain wall fractal dimension for bimodal disorder,
as predicted by droplet theory, is not a well defined quantity.
The reason for this is that there exists
no clear prescription for its estimation if the
domain wall entropy distributions are significantly far from normal.
A brief overview of the method is given in Sec. \ref{sec:method}.
This is followed by our results in Sec. \ref{sec:results} and a brief
discussion in Sec. \ref{sec:discuss}.

\section{\label{sec:method}Background}

The planar Ising model is known to be isomorphic to a system of noninteracting
fermions. One particular illustration \cite {GH64} has been adapted by
Blackman \cite {B82} for disordered systems. For the square lattice, each site
is decorated with four fermions. Equivalently, we can decorate each bond with
two fermions. For a system of $N$ lattice sites, we have $4N$ fermions in total.
It is useful to think of the two fermions decorating a bond to be placed one on
either side. In this way a plaquette (square) is decorated with four fermions;
left, right, top and bottom. 

The partition function for the Ising model on a square lattice with any
set of exchange interactions takes the form

\begin{equation}
Z=2^{N}[\prod_{<ij>} \cosh (J_{ij}/kT)] \hspace {2 mm} (\det D)^{1/2}
\label{e:Z}
\end{equation}
where the product is over all nearest neighbor bonds $J_{ij}$ on
the $N$ site lattice and $D$ is a skew-symmetric ($4N \times 4N$)
matrix. The square root of the determinant of $D$ is also called the
Pfaffian. Essentially, it represents the sum over all closed lattice
polygons and is equal to the product of all the positive eigenvalues
of $D$.

The calculation of the partition function with bimodal disorder has been
described in much detail previously \cite {BP91} and a simple summary
should suffice here. At zero
temperature, $D$ is a singular matrix with a set of degenerate zero
eigenvalues exactly equal in number to the total number of frustrated
plaquettes. In order to extract the physics of the system, degenerate state
perturbation theory is applied at an arbitrarily low temperature. The
defect eigenvalues occur in pairs and approach zero as some power of
$\exp(-2J/kT)$

\begin{equation}
\epsilon=\pm \frac{1}{2}X \exp(-2Jr/kT)
 \label{e:eps}
\end{equation}
where $r$ is an integer (an order of perturbation theory) and $X$ is a real
number that is independent of temperature and depends only on the configuration
of frustrated plaquettes. The ground state energy and entropy can be
expressed exactly as

\begin{equation}
E=-2NJ+2J \sum_{d} r_d
 \label{e:deltaf}
\end{equation}

\begin{equation}
S=k \sum_{d} \ln X_d
 \label{e:s}
\end{equation}
where the sums are over all defect eigenstate pairs.

To summarise the perturbation theory, we first write the matrix $D$ exactly as the
sum of two terms $D=D_{0}+\delta D_{1}$,
where $\delta=1-t$ with $t=\tanh(J/kT)$. Of course $t=1$ and $\delta=0$
in the ground state. Both of the matrices $D_{0}$ and $D_{1}$ are independent
of temperature. The matrix $D_{0}$ has eigenvectors localised inside each
frustrated plaquette; expanded in the basis of the four decorating fermions.
It is these localised states that form the defect basis for the perturbation
theory. The matrix $D_{1}$ is $2 \times 2$ block diagonal in the pairs of
fermions decorating the bonds (one fermion either side). All degeneracy at
first order is lifted by diagonalizing $D_{1}$ in the defect basis. For example,
we can think of just two neighboring frustrated plaquettes. The perturbation
theory gives one defect pair with $r_d =1$ and $X_d = 1$.

In general, the first order calculation will leave some zero eigenvalues. The
corresponding eigenvectors of $D_{1}$ form the basis for second order. We can
imagine a system of two next nearest neighbor frustrated plaquettes.
Clearly $r_d =2$ and $X_d = 1$ or $2$ depending on the arrangement.
In order to show this we need to use the continuum Green's function
\cite {BP91,PB05} $g_{c} = g_{c1} + g_{c2}$. The matrix $g_{c1}$ is
$4 \times 4$ block diagonal in the four fermions inside
each plaquette and clearly allows us to connect two frustrated plaquettes,
across an unfrustrated plaquette. The second order calculation
is performed by diagonalizing $D_{2} = D_{1} g_{c1} D_{1}$. The matrix
$g_{c2}$ is, just like $D_{1}$, $2 \times 2$ block diagonal in the pairs of
fermions decorating bonds. A proof that $g_{c2}$ is irrelevant for the
ground state has been given in Ref. \onlinecite {PB05}.

For higher orders we require Green's functions for states
whose degeneracy has already been lifted. We define, for $r \geq 1$,

\begin{equation}
G_r = -\sum_{i=1}^{N(r)} \mid r,i \rangle (1/\epsilon_r^i) \langle
r,i \mid
\label{e:GR}
\end{equation}
where $\mid r,i \rangle $ denotes state $i$
(with eigenvalue $\epsilon_r^i$) in the set of states whose
degeneracy was lifted at order $r$; there are $N(r)$ of these states.
At third order the matrix to be diagonalized is
$D_{3}=D_{2}(1+G_{1}D_{1})g_{c1}D_{1}$
and, generally at arbitrary order
$D_{n}=D_{n-1}(1+G_{n-2}D_{n-2})\cdots(1+G_{1}D_{1})g_{c1}D_{1}$.
The perturbation theory is applied order by order until all
degeneracy is lifted.

The scheme outlined above allows exact calculations of energy and entropy
in the ground state. The method is fully gauge invariant in that it depends
only on the number and distribution of frustrated plaquettes. Furthermore,
there is no requirement to extrapolate to the ground state.
The Pfaffian is not calculated at any particular numerical value of the
temperature. We believe that this method is the best available for
calculating the ground state entropy of large lattices, although matching
algorithms are better for the energy \cite {HY01}.

We have used periodic boundary conditions in one dimension. The cylindrically
wound frustrated patch was nested in an unfrustrated system of infinite
extent in the second dimension. In this scheme, the introduction of a domain
wall defect is particularly simple. The two plaquettes at the ends of the
defect, one on each side of the patch, change their status; frustrated to
unfrustrated or vice versa. For a perfect ferromagnet this gives a domain
wall energy proportional to $L$ as required. For a fully frustrated system
the defect would make no real difference since the domain wall energy would
be independent of $L$. For the spin glass the domain wall energies are all
multiples of $4J$ since we have $L$ even. The probability that a plaquette
is frustrated is expected to be close to $0.5$ and it is conceivable that
the system with the domain wall could be interchanged with its reference
system in another realization of disorder. This is consistent with the
prediction of droplet theory that the domain wall entropy has random sign.

Since the domain wall entropy is generally a small difference between
two larger numbers, we have taken great care with the floating point
computations. Although our method is analytically exact, it is subject
to numerical propagation error on the computer. Ill-conditioned disorder
realizations were detected by calculating the correlation function along
a path around the cylinder and repeated in arbitrary precision arithmetic
as necessary.

\section{\label{sec:results}Results}

We have calculated the domain wall entropy for bimodal disorder on 
$L \times L$ lattices where $L=8$, $12$, $16$, $24$, $32$, $48$, $64$,
$96$, $128$, $192$ and $256$. For sizes $L \leq 128$, $10^5$ random samples
were taken. We also took $4 \times 10^4$ for $L=192$ and $10^4$ for $L=256$.

\begin{figure}[h]
\includegraphics*[angle=-90,width=8.5cm]{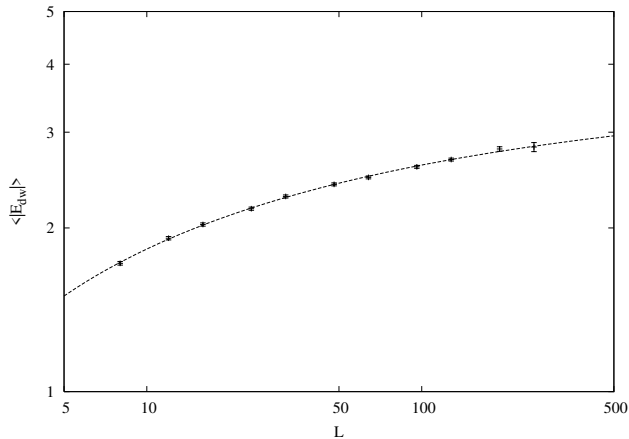}
\caption{\label{f:Fig1} The domain wall energy (in units of $J$) as a function
of system size $L$. The error bars are two standard deviations of the mean and
the curve is a best fit to the form $A-B L^{-p}$ following
Ref. \onlinecite {HY01}.}
\end{figure}

To establish the credentials of our boundary scheme, we have calculated the
domain wall energy. The data is shown in Fig. \ref{f:Fig1} where the error
bars are two standard deviations of the mean. Following
Ref. \onlinecite {HY01}, we have fitted the data to a function $A - B L^{-p}$
and find $A=3.7(2)$, $B=3.2(1)$ and $p=0.23(4)$. We note that the fit is
approaching saturation from below. We believe that this is as a consequence 
of our boundary conditions. The probability of a zero energy ($E_{dw} = 0$)
domain wall is found to decrease with $L$, contrary to the situation with free boundary
conditions in the unwound dimension \cite {FH07}. Furthermore, since we only use even values
of $L$, the defect energies are all multiples of $4J$.
The quality of the nonlinear fit \cite {numrecip} is $Q=0.91$.
Attempts to fit a power law were not successful. For instance a fit
for $L \geq 96$ completely missed the point at $L=256$. We conclude that our
method is reliable although larger system sizes $L>256$ are required to
conclude more convincingly that $\theta=0$.

\begin{figure}[h]
\includegraphics*[angle=-90,width=8.5cm]{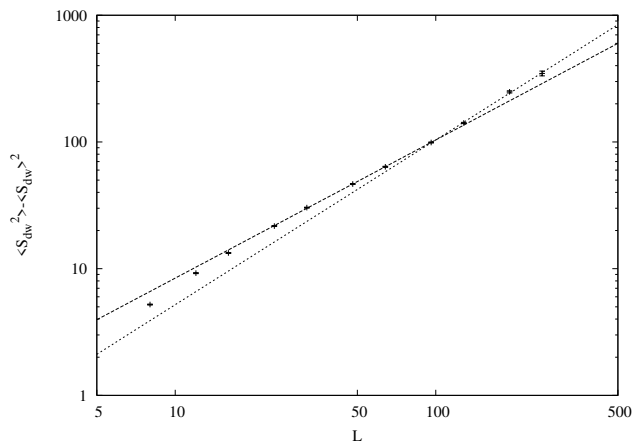}
\caption{\label{f:Fig2} The variance of the domain wall entropy as a function
of system size $L$. The error bars are two standard deviations of the mean.
Two power law fits are shown; for $24 \leq L \leq 96$ (dashed line) and
$L \geq 96$ (dotted line). The powers are respectively $1.090(8)$ and
$1.30(3)$.}
\end{figure}

The variance of the domain wall entropy $S_{dw}$ is shown in Fig.
\ref {f:Fig2}. A power law fit for $24 \leq L \leq 96$ predicts that,
according to Eq. (\ref{e:variance}),
the fractal dimension of the domain walls is
$d_f = 1.090(8)$ where the quality of the fit is $Q=0.42$.
A similar fit ($Q = 0.30$) of $<|S_{dw}|>^2$ gives $1.080(9)$ which agrees
well, indicating good quality data distributions.

A second power law fit for $L \geq 96$, also shown in Fig. \ref {f:Fig2},
reveals a significantly higher value $d_f = 1.30(3)$ with $Q = 0.16$.
However, a fit of $<|S_{dw}|>^2$ gives only $1.23(2)$. Although the quality
$Q = 0.05$ is lower, the difference is too large to be disregarded.
The reason for this discrepancy must lie in the quality
of the distributions for $S_{dw}$. For $L = 256$, for example, the distribution
has skewness $0.64$ and kurtosis $2.06$. Although the mean $<S_{dw}> = 3.53$ is
still much less than the variance ($ \approx 350$) it reflects a
significant sign disparity. The bias most likely arises due to
correlations in the distribution of frustrated plaquettes.
Incidently, the corresponding distributions for $E_{dw}$ are of excellent
quality. For $L = 256$, the skewness and kurtosis are respectively
$-0.006$ and $0.07$.

\section{\label{sec:discuss}Discussion}

In summary, the distributions of the domain wall entropy for large $L$, with bimodal 
disorder in two dimensions,
are found to deviate significantly from normal. In consequence, even if we assume that
droplet theory \cite {FH88} is appropriate, it is unable to prescribe exactly how to get
the domain wall fractal dimension. This does not necessarily mean that droplet theory
is entirely wrong. It just does not give the whole story, only an approximation, for this system;
having a large ground state
degeneracy, a zero critical temperature and a zero stiffness exponent.
Of course, it is quite probable that corrections to scaling are large and difficult to
manage. This is actually a rather likely scenario in view of the poor results for fitting the
ground state energy with cylindrical winding in one dimension \cite {CHK04}.
Scaling corrections are an issue probably related to strong correlations in
the distribution of frustrated plaquettes for large $L$. All gauge invariant quantities like entropy
depend only on the frustrated plaquette distribution; nothing else. The prediction of droplet theory that
the domain wall entropy is normally distributed with zero, or very small, mean 
and large variance probably relates to an assumption that a defect system occurs as a
reference system in another realization of disorder. This assumption may not be true
if the frustrated plaquette distributions are strongly correlated, as is likely in view of the
anamolous behaviour of the degeneracy of the first excitations mentioned earlier.
A further scenario is that due to Hartmann \cite {Hart07}, which proposes that droplet
theory is actually inappropriate for estimating the stiffness exponent of domain wall defects.
If this is correct, it is also unlikely that the fractal dimension of domain walls has 
anything to do with droplet theory. Nevertheless, our results do indicate an approximate
appropriateness for droplet theory in the sense that all the possible estimates for $d_f$
do not seem unreasonable.

Previous studies of the domain wall entropy have worked with much smaller
system sizes. Saul and Kardar \cite {SK93} had sizes up to
$L=36$ and found $d_f \approx 1.0$, while Lukic {\it et al} \cite {LMM06}
used sizes up to $L=50$ and fitted $<|S_{dw}|>$ to find $d_f = 1.03(2)$.
Fisch \cite {Fisch} used sizes up to $L=48$ and has introduced the
idea that the domain wall entropy may be significantly correlated with
energy. It is argued that an effective $d_f$ increases as a function of
$|E_{dw}|$ and convergence to the value $1.25$ consistent with
Eq. (\ref{e:SLE}) is not ruled out. We have tested these predictions and find
that, for $E_{dw} = 0$, the response is in fact much stronger in both the
intermediate and large size regimes. For $L \geq 96$ we find
$d_f = 1.43(4)$ while, for $E_{dw} \not= 0$, $d_f = 1.22(3)$.
Also, the probability of finding a disorder realization with $E_{dw}=0$ is
under $0.5$ for $L > 24$, much less than $0.75$. The cause of these
discrepancies is most probably due to boundary conditions and system size.
We do not have any evidence from our work that the particular values of
$E_{dw}$ are significant for the droplet theory.

We also note that a topological analysis of ground states \cite {RRR07}
predicts $d_f = 1.30(1)$. Essentially, the technique measured the average
length of domain walls. These lengths respond faster than just $L$ since the
domain walls bend to avoid the rigid lattice. Nevertheless, only one ground
state configuration was studied for each disorder realization, completely
ignoring the entropy issue. A study in a similar vein \cite {MH07} looks at the 
properties of minimum energy domain walls and places the fractal dimension
in an interval $1.095(1) \leq d_f \leq 1.395(1)$. This may agree to some 
extent with the point that it is not possible, or very difficult, to actually
pin down the value of $d_f$.

\begin{acknowledgments}
We would like to thank A. K. Hartmann for an enlightening correspondence.
A. Aromsawa acknowledges support from the Thailand Research Fund in the
form of a scholarship from the Royal Golden Jubilee Ph.D. Programme.
\end{acknowledgments}

\end{document}